\documentclass[a4paper,11pt]{article}
\pdfoutput=1 % if your are submitting a pdflatex (i.e. if you have
\usepackage{jcappub} % for details on the use of the package, please
\usepackage[T1]{fontenc} % if needed

\usepackage{subcaption}
\usepackage{siunitx}
\usepackage{hyperref}
\usepackage{braket}
\usepackage{comment}
\usepackage{slashed}
\usepackage{dsfont}
\usepackage{booktabs}
\usepackage{caption} % Allows footnotes in figure captions
\usepackage{multirow}

\definecolor{red}{rgb}{0.9, 0,0}
\definecolor{cerulean}{rgb}{0., 0.62,0.9}
\definecolor{navy}{rgb}{0.05, 0.05,0.8}
\definecolor{orange}{rgb}{0.8, 0.4, 0.}

\renewcommand{\eqref}[1]{Eq.~(\ref{#1})}

\definecolor{pastelBrown}{RGB}{181, 150, 114}

\newcommand{\vrel}{v_{\mathrm{rel}}}

\definecolor{red}{rgb}{0.9, 0,0}
\definecolor{cerulean}{rgb}{0., 0.62,0.9}
\definecolor{navy}{rgb}{0.05, 0.05,0.8}
\definecolor{orange}{rgb}{0.8, 0.4, 0.}

\renewcommand{\eqref}[1]{Eq.~(\ref{#1})}

\title{Minimal Dark Matter in the sky: \\updated Indirect Detection probes}

\author[a,b]{Mohammad Aghaie}
\author[d]{Alessandro Dondarini}
\author[a,b]{Giulio Marino}
\author[a,b]{Paolo Panci}

\affiliation[a]{Dipartimento di Fisica E. Fermi, Universit\`a di Pisa, Largo B. Pontecorvo 3, I-56127 Pisa, Italy}
\affiliation[b]{INFN, Sezione di Pisa, Largo Bruno Pontecorvo 3, I-56127 Pisa, Italy}
\affiliation[d]{Galileo Galilei Institute for Theoretical Physics, Largo Enrico Fermi 2, I-50125 Firenze, Italy}

\emailAdd{mohammad.aghaiemoghadamozbak@phd.unipi.it}
\emailAdd{alessandro.dondarini@phd.unipi.it}
\emailAdd{giulio.marino@phd.unipi.it}
\emailAdd{paolo.panci@unipi.it}

\abstract{
Minimal Dark Matter is among the simplest and most predictive Dark Matter frameworks, with the Majorana SU(2) 5-plet as its smallest accidentally stable real representation. We present a comprehensive reassessment of its indirect-detection signals. The $\gamma$-ray flux from both Sommerfeld-enhanced annihilations and bound-state formation is calculated, incorporating next-to-leading-order corrections and next-to-leading-log resummation of the relevant electroweak effects. In the Milky Way halo, bound-state formation dominates the flux near 100 GeV. The corresponding low-energy spectrum is used to place constraints based on \textsc{Fermi-LAT} observations of Galactic diffuse emission, while the high-energy part of the spectrum is employed to forecast the required observation time for several of the Milky Way's dwarf spheroidal galaxies using the  Cherenkov Telescope Array Observatory (CTAO). \textsc{Fermi-LAT} data strongly disfavor the lower edge of the thermal mass window, even under conservative assumptions about the inner Galaxy density profile. Furthermore, several hundred hours of forthcoming \textsc{CTAO} observations of northern dwarfs should be sufficient to probe the central mass value.
}

\begin{document}
\maketitle
\flushbottom
\section{Introduction}
%%%%%%%%%%%%%%%%%%%%%%%%%%%%%%%%%%%%%%%%%%%%%%%%%%%%%%%%%%%%%%%%%%%%%%

The discovery of the Higgs boson with a mass of 125.1 GeV at the Large Hadron Collider (LHC) represents a major success for the Standard Model (SM). However, the LHC has not found convincing evidence for new Physics at the TeV scale, which was expected to explain the lightness of the Higgs boson's mass. This casts doubts on the relevance of the naturalness concept and the importance of this scale in modern particle physics.

Conversely, the possibility that Dark Matter (DM) consists of a new Weakly Interacting Massive Particle (WIMP), thermally produced in the early Universe, remains a primary motivation for exploring new physics in the multi-TeV mass range. Regardless of its connection to the naturalness problem, a particle within this mass range and with weak SM interactions can naturally explain the observed DM abundance through the thermal freeze-out mechanism, which is a notable accomplishment. Even if this hypothesis is eventually disproven, it serves as a well-founded guiding principle in the quest to identify the particle responsible for the observed DM abundance.

Betting on the possibility of DM being a pure WIMP and the SM holding as the ultimate theory up to the Grand Unified Theory or Planck scale, the Minimal Dark Matter (MDM) model was introduced in 2005~\cite{Cirelli_2006, Cirelli_2007, Cirelli:2008id, Cirelli_2009}, with subsequent generalized versions discussed in~\cite{DelNobile:2015bqo, Bottaro:2021snn, Panci:2024oqc, Bottaro:2022one}. Briefly, this approach involves adding only one extra electroweak (EW) multiplet, $\chi$, to the SM. The goal is to determine the minimal quantum numbers (spin, isospin $I$, and hypercharge $Y$) for $\chi$ that make it a viable DM candidate without compromising the strengths of the SM. No arbitrary features are introduced; the stability of the resulting candidates is ensured by the SM gauge symmetry and renormalizability. By adhering to these principles of consistency and applying the most evident phenomenological constraints (such as direct detection bounds from scattering on nuclei), the theory ultimately identifies a real fermionic SU(2) 5-plet with $Y=0$ as the only candidate that, through its electrically neutral component $\chi_0$, provides a viable DM candidate\footnote{The lightest particle in any EW representation can be made stable by enforcing a symmetry acting solely on the DM. For multiplets with $n \geq 5$, such a symmetry occurs accidentally in the renormalizable Lagrangian. However, for $n>5$, DM stability requires additional assumptions about the theory's UV completion, making the real fermionic 5-plet a special case.}.  The annihilation cross sections of EW multiplets, incorporating significant non-relativistic and non-perturbative effects such as Sommerfeld enhancement (SE)~\cite{Hisano:2003ec, Hisano:2006nn, Arkani_Hamed_2009, Bottaro:2023wjv, Blum:2016nrz} and DM bound state (BS) formation~\cite{Mitridate:2017izz, Bottaro:2021snn} (see Refs.~\cite{Cirelli:2016rnw, Harz:2018csl} for earlier
computations in other contexts), are calculable within the framework of EW theory. Solving the Boltzmann equation to match the observed relic abundance determines the DM candidate's mass to be approximately $M_{\rm DM}=13.7^{+0.6}_{-0.3}$ TeV~\cite{Bloch:2024suj, Bottaro:2023wjv}. One-loop EW corrections split the multiplet components, leading the charged components to decay rapidly into the lightest one, $\chi_0$, which becomes the cosmological DM. The theory's minimality ensures there are no free parameters, enabling precise predictions of its phenomenological signatures.

Nearly twenty years later, within the context described above, significant advancements have been made in computing the thermal mass and main observables. It is pertinent to assess whether the MDM 5-plet still holds up against current data. Generally, its high predicted mass results in a severely suppressed production cross section at the LHC and prospects for future colliders (see e.g.~\cite{Cirelli:2014dsa, Ostdiek:2015aga, Bottaro:2021snn}). The sensitivities of current DM direct detection experiments~(see e.g.~\cite{Hisano:2011cs, Cirelli:2013ufw}) are far from the expected number of events; only the next generation of kiloton-scale experiments may achieve the necessary sensitivity~\cite{Bloch:2024suj,Hisano:2004pv,Hisano:2015rsa, Hisano:2011cs}. On the other hand, indirect detection (ID), initially explored in~\cite{Cirelli:2015bda,Cirelli:2007xd,Cirelli:2008id}, remains the most promising strategy.  

\smallskip
In this study, we focus on $\gamma$-rays as the primary messengers for ID, motivated by the distinctive and highly predictable spectral features of the 5-plet. These features, which span the entire photon-flux spectrum and arise from non-perturbative effects, make the 5-plet and EW multiplets in general compelling targets for multi-TeV $\gamma$-ray astronomy. This field is entering a promising era with the advent of next-generation facilities, led by the Cherenkov Telescope Array Observatory (CTAO)~\cite{CTAConsortium:2010umy, CTAConsortium:2017dvg}. CTAO will operate two sites to provide full-sky coverage: CTAO-South~\cite{ctao_south} at Paranal in Chile and CTAO-North~\cite{ctao_north} at the Roque de los Muchachos Observatory on La Palma in the Canary Islands. CTAO-South, optimized for observations of the Galactic Center (GC) due to its low zenith angles, is expected to reach its full “Alpha” configuration by 2031. CTAO-North  is already partially operational: its first Large-Sized Telescope (LST) has achieved first light, and a four-LST sub-array is anticipated to be fully functional by 2026. While construction progresses in the south, it is therefore strategic to allocate observing time at the northern site to dwarf spheroidal (dSph) galaxies, which offer clean and complementary targets for DM searches.

\smallskip
We extend previous analyses in several directions. First, we recompute the Sommerfeld-enhanced annihilation cross sections into all EW gauge bosons, incorporating next-to-leading-order (NLO) corrections to the non-relativistic EW potential. We next evaluate the DM bound-state-formation (BSF) cross section as a function of the relative velocity of the incoming particles, finding that the dominant bound state exhibits a \textit{p}-wave velocity dependence. With these cross sections in hand, we derive the complete $\gamma$-ray spectrum, resumming NLL contribution that are relevant at the end-point of the spectrum. Our Galaxy emerges as the optimal target: in the Milky Way (MW) halo, BSF enhances the flux near $100\;\text{GeV}$ with a sharp feature in the $\gamma$-ray flux. We use this low-energy part to update previous constraints from \textsc{Fermi-LAT} observations of Galactic diffuse emission~\cite{Cirelli:2015bda}. By contrast, the high-energy part, only mildly affected by BSF, serves to forecast the sensitivity of the CTAO to several northern dSphs, widely regarded as the cleanest laboratories for indirect DM searches. 

\smallskip
The paper is organized as follows. Section~\ref{sec:stage} outlines the theoretical framework. Sec.~\ref{sec:GammaRay} compares the various components of the $\gamma$-ray spectrum across the full energy range. Sec.~\ref{sec:FermiConstr} concentrates on the low-energy regime, updating \textsc{Fermi-LAT} constraints and assessing the impact on the uncertainty of the DM density profile. Sec.~\ref{sec:CTAConstr} examines the high-energy regime, presenting a detailed CTAO sensitivity study, providing the observation time required to exclude the 5-plet at the 95\% confidence level. Finally, Sec.~\ref{sec:conclusions} summarises our main findings.

%%%%%%%%%%%%%%%%%%%%%%%%%%%%%%%%%%%%%%%%%%%%%%%%%%%%%%%%%%%%%%%%%%%%%%
\section{Setting the Stage}
\label{sec:stage}

In this section, we present the theoretical framework of the MDM SU(2) 5-plet. In particular we discuss the impact of NLO corrections on the Sommerfeld-enhanced annihilation, BSF and thermal relic mass.

The model under consideration extends the SM by introducing a single EW Majorana 5-plet $\chi$ with $Y=0$. The general renormalizable Lagrangian is given by  
\begin{equation}
\mathcal{L} = \mathcal{L}_{\rm SM} + \frac{1}{2} \bar{\chi}(i\slashed{D} + M_\chi)\chi \ ,
\label{eq:lag}
\end{equation}
where, $M_\chi$ is the only free parameter of the theory and denotes the mass of the EW multiplet.
The components of the 5-plet, organized by electromagnetic charge, are $\chi = (\chi^{i+}, \chi^0, \chi^{i-})$, with $i$=1 to 2. The electrically neutral component, $\chi^0$, is the lightest state and, by construction, does not couple to the $Z$ boson at tree level. Additionally, EW interactions induce mass splittings between $\chi^0$ and the charged states, given by $\delta_Q = M_{\chi^{Q+}} - M_{\chi^0} = Q^2 \delta_0$, where $\delta_0 = (167 \pm 4)\,\text{MeV}$~\cite{splitting1,splitting2,splitting3}. These splittings allow the charged states to decay into the lightest, stable DM candidate, $\chi^0$. While the mass splittings are negligible at high energies, they play a crucial role in the calculation of Sommerfeld enhancement factors in scattering processes, as well as in the determination of bound-state wavefunctions in low-velocity environments such as the MW's Galactic Halo and dSphs.

\subsection{NLO Sommerfeld Enhancement}
Interactions between charged particles are affected by the Coulomb force when their kinetic energy is low, making the electrostatic potential energy important. In particular, when two $\chi^0$ particles are within a distance of $r\sim m_W^{-1}$, they can exchange virtual EW bosons, causing a potential that significantly perturbs their initial wavefunctions. 
In the context of Feynman diagrams, these interactions are represented by multi-loop ladder diagrams, where mediators are exchanged multiple times before annihilating into SM. While perturbative calculations typically include only a limited number of initial diagrams, Sommerfeld enhancement necessitates the summation of all ladder diagrams, characterizing it as a non-perturbative effect. At low energies, cross sections are primarily governed by s-wave scattering, with p-wave contributions being of lesser significance. 

\smallskip
We focus on the cross section for the process \( n \to \text{SM}\, \text{SM} \), where the initial two-body state, in the low-velocity environoment, is the neutral configuration \( n = 3 \equiv \chi^0 \chi^0 \). We factorize the computation into two steps: \( n \to m \) followed by \( m \to \text{SM}\, \text{SM} \), where \( m \) runs over all states that can mix with the initial state via EW interactions, \textit{i.e.} those sharing the same total charge \( Q \) and spin \( S \). The initial state \( n = 3 \), which has \( Q = 0 \) and \( S = 0 \), mixes with all other neutral two-body states \(\chi^{i+} \chi^{i-}\), namely \( m = \{1, 2, 3\} \equiv \{\chi^{++} \chi^{--}, \chi^+ \chi^-, \chi^0 \chi^0\} \). The dynamics of this coupled system is governed by a potential \( \mathcal{V} \), which can be expressed in matrix form as:
\begin{equation}
    {\cal V}=\begin{pmatrix}
      -4A+8\delta_0 & 2B & 0 \\
       2B & -A+2\delta_0 & 3\sqrt{2}B \\
     0 & 3\sqrt{2}B & 0 \\ 
    \end{pmatrix} 
\end{equation}
where 
\begin{equation}
    \begin{split}
        A&=\alpha/r +c^2_{\rm w}\,\alpha_2/r \, e^{-m_Z\,r}+\alpha_2/r\,\delta V^{{T_3T_3}}_{\rm NLO}(r) \ , \\
        B&=\alpha_2/r \, e^{-m_W\,r}+\alpha_2/r\,\delta V^{W}_{\rm NLO}(r) \ .
    \end{split}
\end{equation}
Here $\alpha$ and $\alpha_{2}$ are the electromagnetic and electroweak fine structure constants, $\alpha=e^2/4\pi$ and $\alpha_{2}=\alpha/s^2_{\rm w}$, and $c_{\rm w}$ and $s_{\rm w}$ are the cosine and sine of the Weinberg angle respectively. The analytic fitting functions $\delta V^W_{\rm NLO}$ and $\delta V^{T_3T_3}_{\rm NLO}$, first derived in \cite{Urban_2021}, take care of the NLO corrections in the spontaneously broken phase of the SM accounting for all possible one-loop topologies correcting the tree-level gauge bosons exchange. We stress that their applicability holds if $M_\chi \gg m_Z$.

\begin{figure*}[t!]
\begin{center}
    \includegraphics[width=0.6\textwidth]{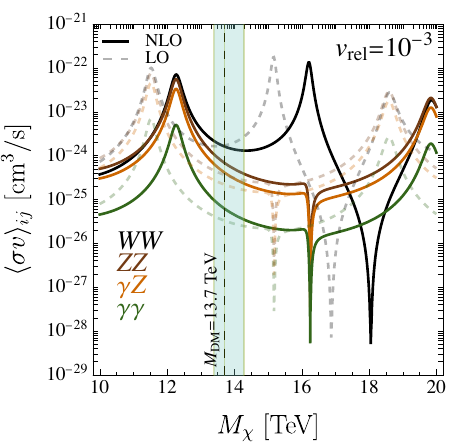} 
	   	\caption{SE annihilation cross sections into EW gauge boson channels, namely $WW$ (black), $ZZ$ (brown), $\gamma Z$ (orange), and $\gamma\gamma$ (green), are shown as a function of the DM mass for a fixed relative velocity $\vrel = 10^{-3}$. Solid lines include NLO corrections, while faint dashed lines correspond to the LO results. The aquamarine band denotes the thermal mass window, namely $13.7^{+0.7}_{-0.3}~\si{TeV}$.}
		\label{fig:annxsecs}
\end{center}
\end{figure*}

The annihilation cross section including the Sommerfeld corrections can be written as
\begin{equation}\label{eq:sigmav3}
    (\sigma v)_3 = 2 s^*_{3m}s_{3n} \Gamma_{mn} \ , 
\end{equation}
where the factor 2 arises from the Majorana nature of the $\chi^0\chi^0$ state and $s$ is the Sommerfeld matrix. 

Here the  tree-level rate matrix $\Gamma_{mn}$ is expressed as:
\begin{equation}
\Gamma_{mn} = \frac{\pi \alpha_2^2}{M_\chi^2} \left[ 
\begin{pmatrix}
2 & -5 & 3\sqrt{2} \\
-5 & \frac{25}{2} & -\frac{15}{\sqrt{2}} \\
3\sqrt{2} & -\frac{15}{\sqrt{2}} & 9
\end{pmatrix} +
\begin{pmatrix}
16 & -4 & 0 \\
-4 & 1 & 0 \\
0 & 0 & 0
\end{pmatrix} \right]\ ,
\end{equation}
where the first matrix refers to the cross section into $WW$, and the second one to the cross sections into all the other gauge bosons. The cross sections into $ZZ$, $\gamma\gamma$ and $\gamma Z$ can be obtained from that second matrix via multiplication by $c_{\rm w}^4$,  $s_{\rm w}^4$ and  $1-s_{\rm w}^4- c_{\rm w}^4$ respectively.
The Sommerfeld factors $s_{3m}$ encode the non-perturbative dynamics arising in the non-relativistic regime and are derived by solving the following system
\begin{equation}\label{eq:schr_sys}
\left\{
    \begin{split}
        &-\frac{1}{M_\chi}\frac{d^2\psi_n^{(m)}(r)}{dr^2}+\mathcal{V}_{ns}(r)\psi_s^{(m)}(r)=M_\chi \beta^2\psi_n^{(m)}(r)\\
        &\psi_n^{(m)}(0)=\delta_{nm},\\ &\frac{d\psi_n^{(m)}}{dr}(r\rightarrow\infty)=i M_\chi \beta \sqrt{1-\frac{\delta_n}{M_\chi \beta^2}}\psi_n^{(m)}(r)\ ,
    \end{split}
    \right.
\end{equation}

where $\phi_n^{(m)}(\vec{r})=r\psi_n^{(m)}(r)/\sqrt{4\pi}$ denotes the reduced wave-function. Here, $\beta = \vrel/2$ where $\vrel$ is the relative velocity of the initial DM particles.  Given~\eqref{eq:schr_sys}, we can read the Sommerfeld factors from the asymptotic behavior of the solution $\psi_3^{(m)}(r\rightarrow \infty)=s_{3m}e^{iM_\chi v r/2}$. For the typical DM velocities in the MW and dSphs galaxies, $\delta_i>M_\chi \beta^2$, which implies that EW interactions cannot convert the initial $\chi^0\chi^0$ state into a pair on-shell charged states. Under this condition, following \cite{Garcia_Cely_2015}, it is convenient to introduce the matrices $g_{im}(r)=\psi_i^{(m)}(r)$ and $h(r)=g'(r)g^{-1}(r)$. In terms of $h(r)$ the Schr\"odinger equations can be reformulated as a set of first-order differential equations
\begin{equation}
    h'(r) + h(r)^2 + M_\chi(M_\chi\beta^2\mathds{1} - {\cal V}(r)) = 0 \ ,
\end{equation}
with boundary condition given by 
\begin{equation}
    h(r\rightarrow \infty) = i \beta M_\chi \sqrt{\mathds{1} - \frac{{\cal V}(\infty)}{M_\chi^2 \beta^2}} \ .
\end{equation}
The key advantage of this algorithm is that, instead of solving a second-order differential equation with boundary conditions at two different points, the problem reduces to solving a first-order differential equation with a single boundary condition at infinity. Consequently, the SE factors are expressed as:
\begin{equation}
    s_{3m}^*s_{3n} = \frac{1}{2iM_\chi\beta}(h_{mn}(0) - h_{mn}^\dagger(0)) \ .
\end{equation}

\smallskip
At larger velocities, such that $\delta_i<M_\chi\beta^2$, the algorithm introduced in \cite{Garcia_Cely_2015} no longer applies. In this case, we solve the Schroedinger system in \eqref{eq:schr_sys} by using the so-called Variable Phase Method (VPM), discussed in detail in \cite{Asadi:2016ybp}.

\subsection{Bound State Formation }
\label{sec:BSF}
The same long-range potential that gives rise to SE is also responsible for BSF. The capture of the initial two-body system occurs at leading order via the emission of a single EW gauge boson. During this process, the total orbital angular momentum $L$ and total spin $S$ of the initial system undergo a transition that in the electric dipole approximation is characterized by $\Delta L = 1$ and $\Delta S = 0$. The resulting bound state is unstable and eventually annihilates into SM, thus contributing to the $\gamma$-ray spectrum from DM annihilations. To characterize the spectrum resulting from BS annihilations, it is useful to decompose the initial $\chi^0\chi^0$ state into eigenstates $|I,I_3\rangle$ of the total isospin $I$:
\begin{equation}
    |\chi^0\chi^0\rangle=\sqrt{\frac{1}{5}}|1,0\rangle-\sqrt{\frac{2}{7}}|5,0\rangle+\sqrt{\frac{18}{35}}|9,0\rangle \ .
\end{equation}

In addition to the selection rules for $L$ and $S$, the electric dipole approximation requires the total isospin to change according to $\Delta I=2$. The potential between two multiplets in a given isospin channel $I$ is attractive whenever $I^2+1-2n^2<0$. This means that, because of the selection rules, the resulting BS from the capture of a $\chi^0\chi^0$ pair is expected to be an almost exact isospin eigenstate with $I=3$, since only channels with $I\leq 5$ are attractive. The spectrum arising from the subsequent annihilation of the BS depends on its total spin $S$. The dominant BS formation channel consists in $p\rightarrow s$ transitions with $S=1$, because bound states with $L\geq 1$ are shallow and their capture cross section suppressed. We will denote BS with $S=1$, principal quantum number $n_B$, zero angular momentum and isospin triplet as $^1(n_Bs)_3$. Once formed, these bound states directly annihilate with a rate $\Gamma_{\rm ann}\sim \alpha_2^5M_\chi$ into SM, in particular into pairs of fermions $f\overline{f}$ and Higgs pair $HH^*$, while the annihilation into pairs of vectors is forbidden by the Landau-Yang theorem. In principle, $^1(n_Bs)_3$ states can also decay to lower lying $^1(n_B'p)_{1,5}$ or $^0(n_B's)_{1,5}$ states. The former states are still an electric dipole transition, again with a rate $\Gamma_{\rm dec}\sim \alpha_2^5M_\chi$, $^1(n_B'p)_{1,5}$, but they annihilate into vectors with a suppressed $\alpha_2^7M_\chi$ rate. Similarly, the decay $^1(n_Bs)_3\rightarrow \,^0(n_Bs)_{1,5}$ is a magnetic dipole transition and as such occurring with a $\alpha_2^7M_\chi$ rate \cite{Bottaro:2021srh}. As a consequence, the BSF cross section is p-wave, which is suppressed in low-velocity environments such as dSphs, and becomes more relevant at higher velocities, as in the MW.

\smallskip
BS give two distinct contributions to the photon annihilation spectrum. First, a series of monochromatic lines at energies given by the bound states binding energies coming from photons emitted directly by the capture process. Second, a contribution to the continuum spectrum due to photons radiated by the fermions and Higgs pair produced by BS annihilations. Both contributions are proportional to the corresponding bound state formation cross section, which have been evaluated following the computations in \cite{Baumgart:2023pwn}. In particular, we have adopted a combination of the variable phase method to determine the initial scattering wave function, and the variational method for the bound states wave functions.

Specifically, the BS with the largest production cross section is \( ^1(1s)_3 \), which for \( M_\chi = 13.7~\si{TeV} \) has a binding energy of \( E_b = 72.94~\si{GeV} \). We find that, within the thermal mass window, the binding energy varies approximately as $E_b/(72.94~\si{GeV})\approx -0.1~\si{GeV}+1.1~\si{GeV}~(M_\chi/13.7~\si{TeV})$. As a consequence, if the \(\gamma\)-ray flux is dominated by BSF, we expect a monochromatic \(\gamma\)-ray line well within the \textsc{Fermi}-LAT energy range, along with a continuum component that must be added to the standard SE annihilation signal.

\begin{figure*}[t!]
\begin{center}
    \includegraphics[width=0.495\textwidth]{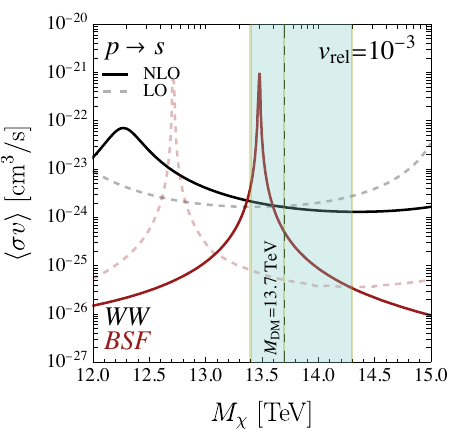} \
    \includegraphics[width=0.485\textwidth]{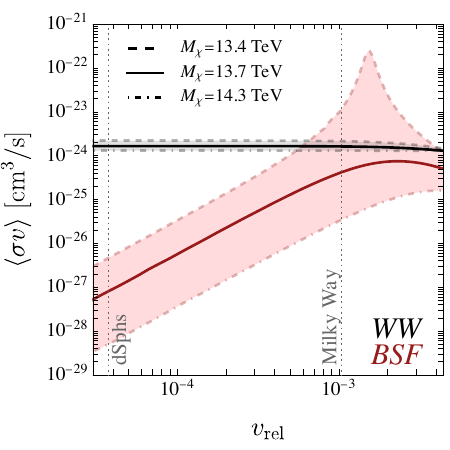} 
    
	   	\caption{
        \textbf{Left panel:} Comparison between the Sommerfeld enhanced annihilation cross section into $WW$ (black lines) and the BSF cross section for the $^1(1s)_3$ bound state formed through a $p \to s $ transition (red lines), both shown as a function of the DM mass for a fixed relative velocity $\vrel = 10^{-3}$. Solid lines include NLO corrections, while faint dashed lines represent the LO results. The aquamarine band denotes the thermal mass window, as in Fig.~\ref{fig:annxsecs}.
        \textbf{Right panel:} Annihilation cross sections for the $WW$ channel (black lines) and BSF (red lines), including NLO corrections, are shown as a function of the relative velocity for three different DM masses within the thermal window: 13.4~TeV (dashed lines), 13.7~TeV (solid lines), and 14.3~TeV (dot-dashed lines). The dotted lines indicate the typical relative velocities: \textit{i.e.} $\vrel=10^{-3}$ for MW and $\vrel\approx 10~\si{km/s}$ for dSphs. The continuum contribution saturates at low velocities, while the BSF cross section, being p-wave, increases with velocity and reaches its maximum for the lowest mass benchmark, around the typical Galactic relative velocity $\vrel= 10^{-3}$. 
}
		\label{fig:BSxsecs}
\end{center}
\end{figure*}

\subsection{Thermal mass}
\label{sec:thermalmass}
The freeze-out prediction of the 5-plet mass was computed in \cite{Mitridate_2017,Bottaro:2021snn}, including both Sommerfeld enhancement and BSF in the annihilation cross section, with further refinement of the theory uncertainty carried out in \cite{Bottaro:2023wjv}. The resulting value for the quintuplet thermal mass is $M_{\chi} = 13.7_{-0.3}^{+0.6}$ TeV, where the theory uncertainty is determined by the approximations assumed in the computation of the annihilation cross section. In particular, since $M_\chi\gg m_W$, the effects of EW symmetry breaking can be neglected until the DM De Broglie wavelength is $\mathcal{O}(m_W)$, that is until $M_\chi v> m_W$ which roughly corresponds to $M_\chi/T \approx 10^4$, $T$ being the temperature of the Universe. At smaller velocities, the shorter range of the Yukawa interactions leads to a saturation of both SE and BSF, except for isolated resonances. As a consequence, we expect the final abundance coming from a proper treatment of the EW symmetry breaking effects to be comprised between the abundance evaluated when $M_\chi/T \approx 10^4$ and that at arbitrarily lower temperatures by extending the SU(2)-symmetric computation well within the broken phase \cite{Bottaro:2021snn}. Another source of uncertainty comes from NLO corrections to the non-relativistic potentials, and thus to SE and BSF \cite{Bottaro:2023wjv}. These NLO corrections include running effects, due to light degrees of freedom flowing in the loop diagrams, and relativistic corrections, in the form of $1/r^2$ and $1/r^3$ terms in the potential. The latter, being relevant at short distances, requires a different matching procedure to extract the SE and the BS wave-functions. 

%%%%%%%%%%%%%%%%%%%%%%%%%%%%%%%%%%%%%%%%%%%%%%%%%%%%%%%%%%%%%%%%%%%%%%

\subsection{Summary}\label{sec:thsumm}
From the theoretical methods and tools discussed above, we can now compute the relevant cross section for low velocity astrophysical environments. 

Before proceeding, let us first specify the relative velocity we adopt in our analysis. For the MW, we consider a truncated Maxwellian distribution with a velocity dispersion $v_0$ and a cutoff at $v = v_{\rm esc}$. We take $v_0 = 220 \,\si{km/s}$ and $v_{\rm esc} = 533~\si{km/s}$, which are typical values for the DM phase-space distribution in our Galaxy (see, e.g.~\cite{Piffl:2013mla}). Assuming an infinite escape velocity, also the two-particle relative velocity distribution is Maxwellian, with $v_{\rm rel} = \sqrt{2} v_0 = 10^{-3}$. For dSphs, we take $v_{\rm rel} \approx 10~\si{km/s}$, based on the typical velocities of stellar tracers in the dwarfs we will discuss later (see~\cite{McConnachie:2012vd}). We use $\langle \sigma v \rangle \equiv \sigma v |_{v=v_{\rm rel}}$ as the velocity-averaged annihilation cross-section. 

In the left panel of Fig.~\ref{fig:annxsecs} we show the annihilation cross sections, from~\eqref{eq:sigmav3}, obtained with (solid curves) and without (dashed curves) NLO corrections. The aquamarine band represents the thermal-mass window (computed at NLO, see Sec.~\ref{sec:thermalmass}), whose central value is marked by a black dashed line at $13.7~\text{TeV}$. Including NLO terms shifts the curves: for the $WW$ channel the cross section is mildly reduced only near the right edge of the thermal-mass window, whereas the cross sections into $\gamma\gamma$, $\gamma Z$, and $ZZ$ increase once NLO corrections are taken into account. Additionally, in the left panel of Fig.~\ref{fig:BSxsecs}, we show the $WW$ annihilation cross section (black lines) as a function of $\vrel$ for three different DM masses: $13.7~\si{TeV}$ (solid line), $13.4~\si{TeV}$ (dashed line), and $14.3~\si{TeV}$ (dot-dashed line). As expected, the SE factor saturates at low velocities, $\vrel \lesssim 10^{-2}$, reaching a value of $\langle\sigma v\rangle_{WW} \approx 10^{-24}~\si{cm^3/s}$. We emphasize that in Fig.~\ref{fig:annxsecs}, the annihilation cross section $\langle \sigma v \rangle_{\gamma \gamma}+\langle \sigma v \rangle_{\gamma Z}/2$ does not directly correspond to the normalization of the flux at the endpoint of the spectrum, as the NLL resummation of the relevant EW effects has not been included, as we will discuss in the following section.

In the right panel of Fig.~\ref{fig:annxsecs}, we compare the BSF cross section of $^1(1s)_3$, computed by fixing the relative velocity to the typical one of the MW (\textit{i.e} $\vrel=10^{-3}$), with the annihilation channel into $WW$, both with (solid lines) and without (faint dashed lines) NLO corrections. Notably, NLO corrections shift the BSF cross section peak into the thermal mass window, yielding a contribution that is comparable, though not dominant, to the Sommerfeld annihilation cross into $WW$ within the thermal mass window. The left panel of Fig.~\ref{fig:BSxsecs} displays the velocity dependence of the BSF cross section (red lines) for three fixed DM masses: \(13.7~\si{TeV}\) (solid), \(13.4~\si{TeV}\) (dashed), and \(14.3~\si{TeV}\) (dot-dashed). As one can see, the \( WW \) channel, and in general the annihilation in all EW gauge bosons, dominates in low-velocity environments, whereas BSF becomes significant in the Galactic halo, potentially leading to observable signatures in \textsc{Fermi}-LAT.

\section{Gamma ray constraints} \label{sec:GammaRay}

In this Section we provide a detailed comparison of the different components of the $\gamma$-ray spectrum coming from annihilating SU(2) 5-plets  with current \textsc{Fermi-LAT} data on the Galactic diffuse emission and evaluate the expected sensitivities of CTAO towards MW's dSphs. 

\medskip
To extract the shape of the photon distribution from the differential cross section, we introduce the photon spectrum ${\rm d} {\cal N}/{\rm d} E_\gamma$ which is normalized with respect to the \textit{line cross section} $\left<\sigma v\right>_{\rm line}^{\rm NLL}$, computed by incorporating both NLO corrections and NLL resummation of the relevant EW effects. Notice that, this cross section is not simply $\left<\sigma v\right>_{\gamma \gamma} + \left<\sigma v\right>_{\gamma Z}/2 $. For the 5-plet, this differs by an $\mathcal{O}(1)$ factor from the NLO result. Further details on the computation of the NLL cross section can be found in Refs.~\cite{Baumgart:2017nsr}. The  photon distribution  can be hence expressed as: 
\begin{equation}
\label{eq:Nspectrum}
    \frac{{\rm d} \cal N}{{\rm d} E_{\gamma}} =  \frac1{\left<\sigma v\right>_{\rm line}^{\rm NLL}} \left( \left. \frac{{\rm d} \langle \sigma v \rangle}{{\rm d} E_\gamma} \right|_{\rm Cont}  \hspace{-.25cm} + \left. \frac{{\rm d} \langle \sigma v \rangle}{{\rm d} E_\gamma} \right|_{\rm BS-line} \hspace{-.45cm} + \left. \frac{{\rm d} \langle \sigma v \rangle}{{\rm d} E_\gamma} \right|_{\rm line}^{\rm NLL} \right) \ .
\end{equation}

\smallskip
The first term represents the continuum contribution, primarily arising from the production of heavy EW gauge bosons in direct annihilation processes. Furthermore, the formation of bound states leads to their prompt decay into SM particles, further enhancing the $\gamma$-ray continuum. Introducing ${\rm d}N^f_\gamma/{\rm d}E_{\gamma}$ as the photon energy spectrum per annihilation into channel $f$, computed using the SM and EW tools from Ref.~\cite{Cirelli:2010xx}, the continuum contribution is given by
\begin{equation}
\label{eq:C_spectrum}
 \left. \frac{{\rm d} \langle \sigma v \rangle}{{\rm d} E_\gamma} \right|_{\rm Cont}  \hspace{-.45cm}  =   
   \langle \sigma v \rangle_{ 1s3}  \hspace{-.2cm}  \sum_{i=\{q,\ell,h\}} \text{BR}_i  \frac{{\rm d} N_\gamma^i}{{\rm d} E_{\gamma}}(M_\chi) +  \langle \sigma v \rangle_{WW}\frac{{\rm d} N_\gamma^{W}}{{\rm d} E_{\gamma}}(M_\chi)+ \frac{\langle \sigma v \rangle_{\rm line}^{\rm NLL}}{\tan^2\theta_w}  \frac{{\rm d} N_\gamma^{Z}}{{\rm d} E_{\gamma}}(M_\chi)  \ ,
\end{equation}
where the velocity-averaged annihilation cross sections are computed in Sec.~\ref{sec:stage}. Here, $\text{BR}_i$ denotes the branching ratio for the annihilation of the $^1(1s)_3$ bound state, which predominantly annihilates into fermions and Higgs bosons via $Z$-mediated $s$-channel diagrams. To avoid double counting hard photons arising from the splitting of the transverse component of the $W$ boson, we switch off the EW corrections associated with this component, as they are already included through the NLL resummation of the endpoint region. Conversely, we retain the EW corrections from the longitudinal component, since they contribute to soft photon emission within the energy window relevant for \textsc{Fermi-LAT}.

\smallskip
The second term of~\eqref{eq:Nspectrum} is given by
\begin{equation}\label{eq:BSline_spectrum}
 \left. \frac{{\rm d} \langle \sigma v \rangle}{{\rm d} E_\gamma} \right|_{\rm BS-line}   \hspace{-.4cm}  =  \frac12 \langle \sigma v \rangle_{ 1s3}   \frac{{\rm d} N_\gamma^\gamma}{{\rm d} E_{\gamma}}(E_b) \ ,
\end{equation} 
and represents the differential rate for the production of a prompt photon due to BSF, centered at the process's binding energy  $E_b$ as discussed in Sec.~\ref{sec:BSF}. If this line emerges from the continuum, it could serve as a potential smoking-gun signature in the $\gamma$-ray spectrum, particularly within the hundreds of GeV energy range. 

The final term of~\eqref{eq:Nspectrum} is
\begin{equation}\label{eq:endline_spectrum}
 \left. \frac{{\rm d} \langle \sigma v \rangle}{{\rm d} E_\gamma} \right|_{\rm line}^{\rm NLL}  \hspace{-.2cm}  =  \langle \sigma v \rangle_{\rm line}^{\rm NLL} \left(2\delta(E_\gamma-M_\chi)+  \mathcal{E}^{\rm NLL}(E_{\gamma})\right) \ ,
\end{equation} 
which represents the differential rate for the production of a hard photon, at the endpoint of the spectrum ($E_\gamma=M_\chi$) 
significantly Sommerfeld enhanced, including NLL resummation of the relevant EW effects.
The reference to hard photon, implies that in addition to the pure line term (the Dirac delta in the parenthesis) we also focus on those final states that carry an order one fraction of the available energy. In particular, defining $z=E_\gamma/M_\chi$, we focus on photons such that $1-z\ll1$. The endpoint contribution $\mathcal{E}^{\rm NLL}(E_\gamma)$ is computed by following Soft Collinear Effective Theory as already discussed in~\cite{Baumgart:2017nsr, Baumgart:2023pwn, Bauer:2000yr, Bauer:2001yt, Bauer:2014ula}.

\begin{figure*}[t!]
\begin{center}
    \includegraphics[width=0.49\textwidth]{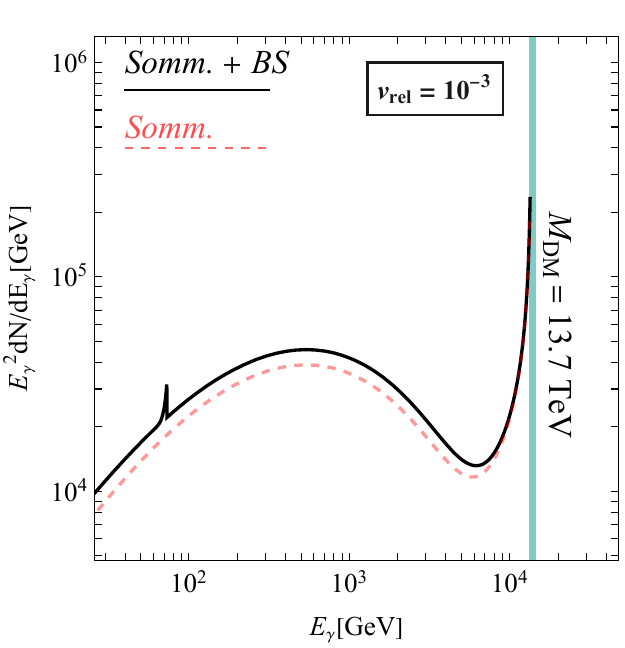} 
    \includegraphics[width=0.49\textwidth]{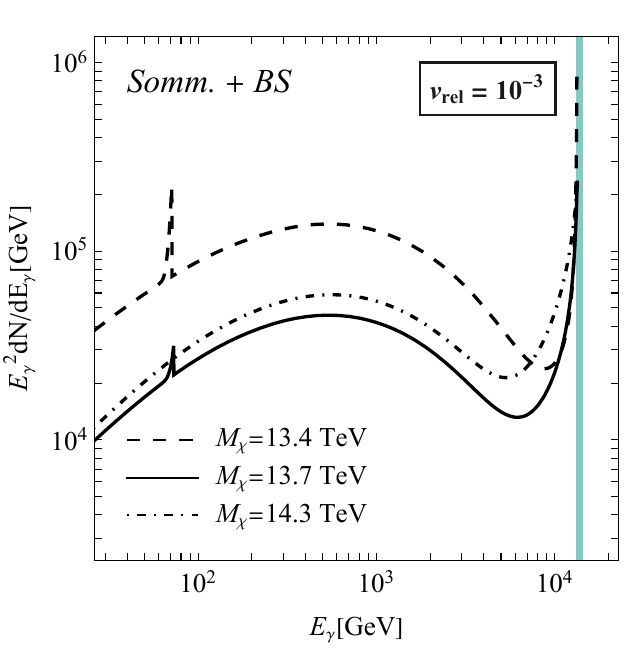}\hfill
    
	   	\caption{Theoretical $\gamma$-ray spectra from the MW halo ($\vrel=10^{-3}$) computed as in~\eqref{eq:Nspectrum}. \textbf{Left Panel}:  Result keeping fixed the central value of the thermal DM window. The red dashed curve represents the SE contribution only, while the black solid line shows the total spectrum, combining both SE and BSF. The aquamarine band correspond to the thermal mass window. \textbf{Right Panel}: Total contribution of the $\gamma$-ray spectrum for three different DM masses within the thermal window: 13.4~TeV (dashed lines), 13.7~TeV (solid lines), and 14.3~TeV (dot-dashed lines).}
		\label{fig:theorySpec}
\end{center}
\end{figure*}

\medskip
 Fig.~\ref{fig:theorySpec} shows the photon energy spectrum, multiplied by $E_{\gamma}^2$, computed for $\vrel=10^{-3}$ which is the typical DM relative velocity in the MW's halo. In the left panel of the figure, the red dashed curve represents the SE contribution, while the black solid line shows the total spectrum, combining SE annihilation and BSF. As evident, two distinct line-like features emerge: the sharp feature peaking near the DM thermal mass originating from~\eqref{eq:endline_spectrum} and a low-energy line peaking to the $^1(1s)_3$ binding energy, $E_{\gamma}\simeq E_{b}\simeq 73$ GeV, coming from~\eqref{eq:BSline_spectrum}. The remaining component of the spectrum is the broad continuum as given in~\eqref{eq:C_spectrum}. As already commented in Sec.~\ref{sec:thsumm}, in dSphs, the BS contribution to the $\gamma$-ray flux is negligible since the BSF cross section is p-wave and therefore suppressed. At this velocity only the Sommerfeld enhancement contribution is sizable and therefore in the CTAO analysis only the red dashed line has to be taken into account.

 In the right panel of the same figure we perform a scan of the  $\gamma$-ray spectrum for three different mass values within the thermal window fixing $\vrel=10^{-3}$. Specifically, we denote as a dashed line the flux for the mass at the lower end of the window, $M_{\chi}=13.4\,\text{TeV}$, with a solid line the flux from the central value $M_{\chi}=13.7\,\text{TeV}$ and with a dot-dashed line the flux arising from the mass value at the higher end of the thermal window, $M_{\chi}=14.3\,\text{TeV}$. As readily seen from the plot, mass values that are closer to the resonant peak of BSF cross section result in a prominent $\gamma$-ray line coming from the radiation of a photon at $E_\gamma\simeq E_b$ accompanied by a continuum.

\medskip
Having at our disposal the photon distribution from annihilating SU(2) 5-plets, the differential prompt flux at Earth from an angular direction ${\rm d}\Omega$ produced by the annihilation of self-conjugated $\chi_0$ is 
\begin{equation}\label{eq:flux}
    \frac{{\rm d}\Phi}{{\rm d} E_{\gamma}{\rm d}\Omega}= \frac{J(\theta)}{8\pi} \frac{\left<\sigma v\right>_{\rm line}^{\rm NLL}}{M_{\chi}^2}  \frac{{\rm d} \cal N}{{\rm d} E_{\gamma}} \ , \qquad \mbox{where } J(\theta) = \int_{\rm l.o.s.} {\rm d} s \, \rho(r(s,\theta))^2  , 
\end{equation}
 is the dimension-full \emph{J-factor} (expressed in $\text{GeV}^2/\text{cm}^5$)  which integrates the intervening matter along the line of sight. In~\eqref{eq:flux}  the galactocentric  and l.o.s. coordinates ($r$ and $s$, respectively) are related as $r(s,\theta)^2=r_{\odot}^2+s^2-2 s r_{\odot} \cos{\theta}$, where $\theta$ is the angle between the line of sight and the axis connecting the Earth to the GC. 

The value of the $J$-factor depends on the choice of the specific target. In the following, we present the selection of astrophysical targets considered in our analysis:

\medskip
{\bf Galactic Halo:} We focus on a specific Region of Interest (RoI) located in the inner part of the MW halo. The optimal RoI is selected based on the following criteria: $i$) the presence of a $\gamma$-ray line at the binding energy of the bound state, which lies well within the \textsc{Fermi-LAT} energy window; $ii)$ a sufficiently large angular size to mitigate uncertainties in the DM density profile near the GC; $iii)$ not so large, however, as to introduce significant mismodelling in the template fit for the Galactic diffuse emission, as further discussed in Sec.~\ref{sec:bkg_extraction}.

\smallskip
Based on these guidelines, we define our fiducial RoI as a circular area centered on the GC of radius $R_{\rm GC}$.  Specifically, following Ref.~\cite{Fermi-LAT:2013thd}, we adopt $R_{\rm GC} = 16^\circ$ (hereafter referred to as RoI16), while masking the Galactic plane in the region defined by $|b| < 5^\circ$ and $|l| > 6^\circ$, where $b$ and $l$ denote Galactic latitude and longitude, respectively. This configuration optimizes the signal-to-noise ratio for a DM distribution following an Einasto profile

\begin{equation}
\label{eq:Ein_profile}
\rho_{\rm Ein}(r) = \rho_s \exp\left\{ -\dfrac{2}{\alpha} \left[ \left( \dfrac{r}{r_s} \right)^\alpha - 1 \right] \right\}.
\end{equation}

We adopt a shape parameter $\alpha = 0.17$ and a scale radius $r_s= 20\,\text{kpc}$ as motivated by Cold Dark Matter numerical simulations~\cite{Navarro:2008kc}. The scale density $\rho_s$ is adjusted such that the local DM density at the position of the sun $r_{\odot}=8.5\,\text{kpc}$ is $\rho(r_{\odot}) = 0.4\,\text{GeV}/\text{cm}^3$~\cite{Catena:2009mf, Salucci:2010qr}. This procedure yields $\rho_s \simeq 0.08\,\text{GeV}/\text{cm}^3$. Using these parameters, and taking into account the masking of the Galactic plane, the resulting $J$-factor is reported in Tab.~\ref{tab:target_MW}.

\medskip
{\bf Dwarf Spheroidal Galaxies:} Dwarf spheroidal galaxies are considered among the most promising environments for DM searches, as they are thought to be DM-dominated with relatively low astrophysical backgrounds. This makes them ideal targets for establishing robust constraints on DM properties. The main uncertainty in analyzing fluxes from dSphs arises from the $J$-factor, which is derived from dynamical mass models using the Jeans equation (for a more detailed discussion on the main uncertainties for the determination of the $J$-factor see e.g. Ref.~\cite{Lefranc:2016dgx}). This uncertainty can be treated as a systematic error in the statistical analysis.

\smallskip

In our study, we focus on a sample of dSphs located in the northern hemisphere. Specifically, following the selection adopted by the CTAO Collaboration~\cite{CTAO:2024wvb}, we consider one \textit{classical} dSph, Draco I (DraI), and three \textit{ultra-faint} dSphs: Willman I (Wil1), Coma Berenices (CBe), and Ursa Major II (UMajII)\footnote{Although the ultra-faint dSph UMajII is not part of the original CTAO selection, we have chosen to include it in our analysis given its favorable sensitivity, as detailed in Sec.~\ref{sec:CTAConstr}.}. While some ultra-faint dSphs may have larger $J$-factors than the classical ones, they also tend to have fewer stellar tracers, which increases the systematic uncertainties in their $J$-factor estimates and can therefore lead to slightly weaker constraints. The $J$-factors for the selected dSphs are provided in Tab.~\ref{tab:target_MW}, where we present their values for a circular region with radius $R_{\rm dSphs} = 0.5^\circ$ centered on each dSph. 

\begin{table}[h!]
    \centering
    \renewcommand{\arraystretch}{1.3}
    \caption{\centering Annihilation $J$-factors for the selected dSphs and the Inner galaxy.}
    \label{tab:target_MW}
    \begin{tabular}{l c c c c c}
        \toprule
        \textbf{Target} 
        & \multicolumn{1}{c}{\textbf{Inner Galaxy}} 
        & \multicolumn{4}{c}{\textbf{dSph (Northern hemisphere)}} \\
        \cmidrule(lr){2-2} \cmidrule(lr){3-6}
        & RoI16 (Einasto)
        & Dra I~\cite{CTAO:2024wvb} 
        & Wil 1~\cite{CTAO:2024wvb} 
        & CBe~\cite{CTAO:2024wvb} 
        & UMajII~\cite{Boddy:2019qak} \\
        \midrule
        $\log_{10}{J}\,\,\,\,[\text{GeV}^2/\text{cm}^5]$ 
        & 23.0
        & $18.7^{+0.3}_{-0.1}$ 
        & $19.1^{+0.6}_{-0.5}$ 
        & $19.5^{+0.9}_{-0.7}$ 
        & $19.44^{+0.41}_{-0.39}$ \\
        \bottomrule
    \end{tabular}
\end{table}

\subsection{Constraints from \textsc{Fermi}-LAT}
\label{sec:FermiConstr}
We now discuss the constraints on the photon spectrum in the hundred-GeV range from 16 years \textsc{Fermi}-LAT data measurements of the galactic diffuse emission.

\subsubsection{Data Selection}\label{sec:fermi_datasel}
In this work we use 830 weeks of data, spanning from August 4, 2008, to July 9, 2024, with energies ranging from 1 GeV to 1 TeV. A zenith-angle cut of $\theta < 100^{\circ}$ is applied to minimize contamination from the Earth's albedo, alongside a quality-filter cut $\textbf{\texttt{DATA\_{QUAL}}} > 0$. Events are selected from good-quality time intervals (GTI), and instances when the \textsc{Fermi-LAT} operated at a rocking angle exceeding \textbf{\texttt{52 degrees}} are excluded\footnote{\textbf{\texttt{LAT\_CONFIG == 1}} \textbf{\texttt{\&\&}} $\textbf{\texttt{ABS(ROCK\_ANGLE)}} < 52^{\circ}$}. To enhance the sensitivity to faint signals, we limit our analysis to \textbf{\texttt{ULTRACLEAN}} (\textbf{\texttt{evclass}} = 512) events and employ the \textbf{\texttt{P8R3\_ULTRACLEAN\_V3}} instrument response functions. The selection of events and exposure map calculations are performed using the 2.2.0 version of \textbf{\texttt{ScienceTools}}.

The data is initially binned into 75 logarithmically-spaced energy bins between 1 GeV and 1 TeV and spatially binned using \textbf{\texttt{HEALPIX}}~\cite{Gorski:2004by} with \textbf{\texttt{nside=512}}. As already stated at the beginning of the section, we follow previous \textsc{Fermi}-LAT collaboration analysis protocols and extract the data for RoI16~\cite{Fermi-LAT:2013thd, Fermi-LAT:2015kyq}. In this region, we mask the Galactic plane with $|b| < 5^\circ$ and $|l| > 6^\circ$, as well as the regions corresponding to 68\% containment around known sources based on the \textbf{\texttt{4FGL\_DR2}} catalog~\cite{Ballet:2020hze}.

\subsubsection{Background Extraction}\label{sec:bkg_extraction}
In our analysis, we model the spectral data within RoI16 as a linear combination of two main components: Galactic diffuse emission and isotropic diffuse emission. Specifically, we employed the \textbf{\texttt{gll\_iem\_v07}}\footnote{\href{https://fermi.gsfc.nasa.gov/ssc/data/access/lat/BackgroundModels.html}{\url{https://fermi.gsfc.nasa.gov/ssc/data/access/lat/BackgroundModels.html}}} model to represent the Galactic diffuse emission. This model was derived using spectral line surveys of H\hspace{.03cm}\textsc{i} and CO~\cite{Fermi4FGLdiffuse2019}, which trace the distribution of interstellar gas and has been refined to account for variations in dust column density. The isotropic diffuse emission, which accounts for unresolved extragalactic contributions and residual charged particle contamination, was modeled using the \textbf{\texttt{iso\_P8R3\_ULTRACLEAN\_V3\_v1}} template\footnote{Note that the \texttt{p8r3} model is not recommended for searches involving extended emission, however, since in our analysis we rely solely on the spectral information of the diffuse model and not its spatial morphology, we are less sensitive to potential mismodeling of the diffuse emission.}.

Additionally, the contributions from the known point sources within our ROI were modeled using the 4FGL catalog\footnote{Even though we are using a few more years of data, yet as it is mentioned in \cite{Gammaldi:2021zdm}, no new sources are found in our ROI. Moreover, the impact of PS is always subleading.}, which provides a comprehensive list of known $\gamma$-ray point sources detected by Fermi-LAT. 

\subsubsection{Statistical Analysis and  Results}\label{sec:statFermi}
Focusing on RoI16 of the inner MW, we perform a spectral analysis of the extracted data. In particular, we restrict to an energy window in the $E_{\gamma} \in \left[52, 398\right]\,\text{GeV}$ range. The lower limit of the energy window is selected to prevent potential mismodelling of the astrophysical background caused by the presence of the  Galactic Center excess~\cite{Goodenough:2009gk}. On the other hand, the upper limit is selected to maximize the sensitivity to the DM signal, as the DM continuum extends into the hundreds of GeV range and the spectral line from BSF is centered at typical binding energies in the $71\,\text{GeV}\lesssim E_{b}\lesssim 76\,\text{GeV}$ range (see Sec.~\ref{sec:BSF}). Furthermore, we have verified that small adjustments to the energy window boundaries have a negligible impact on the results of our analysis. For instance, for the central value $M_{\chi} = 13.7\,\text{TeV}$, extending the upper limit of the energy window to $E_{\gamma} \simeq 525\,\text{GeV}$ by adding three additional bins strengthens the bound by less than 3\%. Similarly, extending the fiducial energy window on the lower end to $E_{\gamma} \simeq 43,\text{GeV}$ by including two additional bins modifies the constraints by less than 8\%. In principle, the analysis could be extended up to 1 TeV; however, at the upper end of the \textsc{Fermi}-LAT energy range, the modeling of the diffuse Galactic emission becomes increasingly uncertain. With the full 16 years of \textsc{Fermi}-LAT data now available, improving the background modeling in this high-energy regime would be highly beneficial. Such an effort could significantly enhance the sensitivity to heavy DM, potentially tightening the bounds from diffuse $\gamma$-ray emission well beyond current limits from the collaboration~\cite{Fermi-LAT:2016zaq}.

\smallskip
The binned signal and background number of events are computed by convolving the corresponding photon flux from the given RoI with the \textit{detector response matrix} (DRM)  and the effective exposure $\mathcal{E}$ (effective area times exposure). In this way, the reconstructed number of events in the $i$th energy bin reads
\begin{equation}\label{eq:fermi_counts}
	N^i_{\rm DM,\, bkg}=\text{DRM}^{i}_{j}\int_{E_j}^{E_{j+1}}\rm{d} E_{\gamma} \frac{d\Phi^{\rm RoI}_{\text{DM,\, bkg}}(E_{\gamma})}{\rm{d}E_{\gamma}}\mathcal{E}^{\rm RoI}(E_{\gamma}) \ .
\end{equation}
Concerning the astrophysical background $d\Phi^{\rm RoI}_{\rm bkg}/dE_{\gamma}$, we use the \textsc{Fermi-LAT} template  for the Galactic diffuse emission combined with the template for isotropic emission, as detailed in Sec.~\ref{sec:bkg_extraction}. In the following, these two components are scaled by two free normalization factors, $A_{\rm diff}$ and $A_{\rm iso}$,  which adjust the overall amplitude of the templates. This step is necessary to tailor the template to the specific data set being used. Such normalizations are considered as nuisance parameters in the analysis, as explained below. On the other hand, the DM number counts are obtained by first integrating~\eqref{eq:flux} over the solid angle of RoI16 and then plugging the resulting $\rm{d}\Phi^{\rm RoI}_{\rm DM}/\rm{d}E_{\gamma}$ in~\eqref{eq:fermi_counts}. Since in the MDM framework there are no free parameters, we scale the DM flux  by introducing a normalization factor $\kappa$, which effectively allows the total annihilation cross section to be treated as a free parameter, $\left<\sigma v\right>\to \kappa \left<\sigma v\right>$, from which we derive the bound.

\begin{figure}[t!]  
	
    \includegraphics[width=0.495\textwidth]{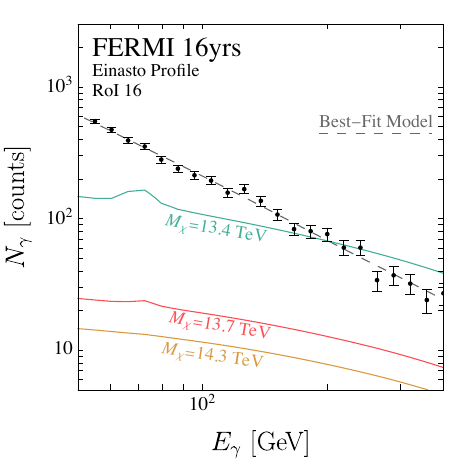} \
    \includegraphics[width=0.475\textwidth]{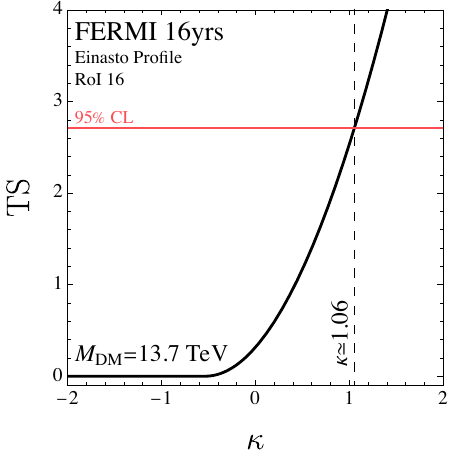}\hfill

	\caption{\textbf{Left panel:} Fit of the photon number counts in RoI16 within the fiducial energy window, using the 16-year Fermi-LAT dataset. The gray dashed line represents the best-fit model, obtained by maximizing the likelihood over the parameter space $\{\kappa, A_{\rm iso}, A_{\rm diff}\}$ for the reference mass $M_{\chi}= 13.7\,\text{TeV}$ (see the text for more details). The brown, red, and blue solid lines show the predicted number of events for three benchmark masses within the thermal mass window, assuming an Einasto density profile (see the discussion on the targets in Sec.~\ref{sec:GammaRay}). Specifically, the brown line corresponds to $M_{\chi} = 14.3\,\text{TeV}$ (the upper end of the thermal window), the red line to $M_{\chi} = 13.7\,\text{TeV}$ (the central value), and the blue line to $M_{\chi} = 13.4\,\text{TeV}$ (the lower end of the thermal window). \textbf{Right panel:} Plot of the TS defined in~\eqref{eq:ts_fermi} as a function of the parameter $\kappa$ for the central mass value $M_{\rm DM}=13.7\,\text{TeV}$. The 95\% one-sided upper limit corresponds approximately to the value of $\kappa$ for which $\mathrm{TS}(\kappa) \approx 2.71$, as denoted by the red solid line.}
	\label{fig:fitRoI16} 
\end{figure}

For a fixed DM mass, we thus build a joint Poissonian likelihood over all the energy bins in the fiducial window
\begin{equation}\label{eq:like_fermi}
	\mathcal{L}(\kappa, A_{\rm iso}, A_{\rm diff})=\prod_{i=1}^{\mathcal{N}}\frac{(N_{\text{th}}^{i}(\kappa, A_{\rm iso},A_{\rm diff}))^{N_{\text{obs}}^{i}}}{N_{\text{obs}}^{i}!}e^{-N_{\text{th}}^{i}(\kappa,  A_{\rm iso}, A_{\rm diff})} \ ,
\end{equation}
where the predicted and observed counts in the $i$th energy bin are $N_{\rm th}^i\equiv \kappa \,N_{\rm DM}^i+A_{\rm iso} N_{\rm iso}^i+A_{\rm diff} N_{\rm diff}^i$ and $N_{\rm obs}^i$. As a first step, we maximize the Poissonian likelihood in~\eqref{eq:like_fermi} in the $\{\kappa, A_{\rm iso},  A_{\rm diff}\}$ parameter space, thus obtaining the best-fit values of the scaling parameters, $\{\hat\kappa, \hat A_{\rm iso},  \hat A_{\rm diff}\}$. 

\smallskip
We present the results of the fit procedure in Fig.~\ref{fig:fitRoI16}. In the left panel, we show the predicted DM counts as a function of the photon energy for three benchmark masses within the thermal window, indicated by solid lines in brown, red, and blue. Specifically, the brown line corresponds to $M_{\chi} = 14.3\,\text{TeV}$ (the upper end of the thermal window), the red line to $M_{\chi} = 13.7\,\text{TeV}$ (the central value), and the blue line to $M_{\chi} = 13.4\,\text{TeV}$ (the lower end). For comparison, we also show  as a gray dashed line the best-fit of the full DM+background model, fixing the reference mass at $M_{\chi} = 13.7\,\text{TeV}$ For this mass, we obtain as best-fit parameters $\{\hat\kappa, \hat A_{\rm iso}, \hat A_{\rm diff}\} \approx \{-0.5, 4.2, 0.9\}$\footnote{Note that, in order to find the \textit{true} maximum of the likelihood, one must allow $\kappa$ to take both positive and negative values.}. We observe that for mass values near the lower edge of the thermal mass window, the spectral feature at low energies has significant constraining power, whereas for masses near the upper edge of the window, the constraining power mainly comes from the data in the high-energy tail. As it is evident, the mass values on the left side of the light blue band are clearly ruled out, whereas for larger masses, a more detailed analysis is required.

 Starting from~\eqref{eq:like_fermi}, we define the test statistic for upper limits $\rm TS(\kappa)$ as
\begin{equation}\label{eq:ts_fermi}
    \begin{split}
        \text{ TS}(\kappa) = 2 \log{\left[\frac{\max_{\kappa, A_{\rm iso}, A_{\rm diff}} \mathcal{L}(\kappa, A_{\rm iso}, A_{\rm diff})}{\max_{A_{\rm iso}, A_{\rm diff}} \mathcal{L}(\kappa, A_{\rm iso}, A_{\rm diff})}\right]} \ ,
    \end{split}
\end{equation}
for $\kappa > \hat{\kappa}$, and zero otherwise. By construction, the numerator in~\eqref{eq:ts_fermi} corresponds to the maximum likelihood evaluated at the best-fit parameters $\{\hat{\kappa}, \hat{A}_{\rm iso}, \hat{A}_{\rm diff}\}$. The denominator represents the profile likelihood, obtained by maximizing the likelihood with respect to the background parameters $A_{\rm iso}$ and $A_{\rm diff}$ for each fixed value of $\kappa$. The TS in~\eqref{eq:ts_fermi} follows an asymptotic $\chi^2$ distribution with one degree of freedom. As a consequence, we can use Wilks' theorem to compute the one-sided 95\% confidence level (CL) upper-limit on $\kappa$ by imposing $\rm TS(\kappa)\simeq 2.71$, see e.g. Ref.~\cite{Rolke:2004mj}. 
\begin{figure*}[t!]
	\centering
	\includegraphics[width=0.49\textwidth]{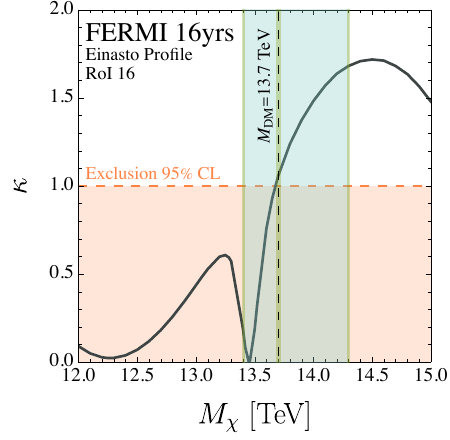} \
    \includegraphics[width=0.49\textwidth]{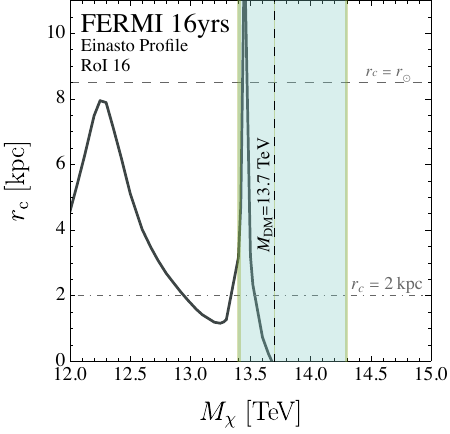}
	\caption{\textbf{Left Panel}: The black solid line shows the 95\% CL upper bound on the rescaling parameter $\kappa$ of the total annihilation cross section as a function of the DM mass $M_{\chi}$ in the $12\,\text{TeV}\le M_{\chi}\le 15\,\text{TeV}$ range. This bound is derived from RoI16 data assuming an Einasto profile. The excluded region, corresponding to $\kappa < 1$, is indicated by the light orange shaded area, while the aquamarine  band denotes the thermal mass window of the MDM 5-plet. \textbf{Right Panel}:  Core size required to weaken the exclusion bound in the left panel to $\bar{\kappa} = 1$, as a function of the DM mass $M_{\chi}$. See the left panel for the color code.
}
	\label{fig:money_k}
\end{figure*}
As an illustrative example, we show in the right panel of Fig.~\ref{fig:fitRoI16} the TS as a function of the rescaling parameter $\kappa$ for the central mass value $M_{\rm DM}$. The red horizontal line indicates the threshold $\mathrm{TS} = 2.71$, corresponding to a 95\% CL upper limit on $\kappa$. For this mass, the upper limit is $\kappa = 1.06$, as marked by the dashed black vertical line.

The results of the aforementioned procedure are summarized in the left panel of Fig.~\ref{fig:money_k}, where we have applied the analysis by varying the DM mass in the $12\,\text{TeV}\le M_{\chi}\le 15\,\text{TeV}$ range. For each mass the candidate is excluded when $\kappa<1$ (light orange area), namely when the confidence interval for the cross section is smaller than the cross section predicted by the model. Importantly, the sharp drop in the exclusion curve at 13.5 TeV is attributed to the peak of the BSF cross section, see the left panel of Fig.~\ref{fig:BSxsecs}. This notable feature allows us to exclude the 5-plet DM candidate  within the lower end of the thermal mass range (denoted by an aquamarine band), since the upper limit obtained from RoI16 using an Einasto profile falls entirely below the exclusion threshold $\bar{\kappa} =1$ in the $13.4\,\text{TeV}\lesssim M_{\chi}\lesssim 13.7\,\text{TeV}$ mass range. For values of $M_{\chi}$ below the thermal mass window, the exclusion curve has another minimum around $M_{\chi} \simeq 12.2\,\text{TeV}$, as this mass value corresponds to the resonant peak of the SE annihilation into $W^+W^-$ (see Fig.~\ref{fig:annxsecs}) resulting in an enhanced continuum signal. Conversely, for masses above the thermal window, the exclusion significantly weakens since such mass values are distant from the resonant peaks of BSF and SE cross sections.

It is important to emphasize that these conclusions depend on the uncertainty in the DM density profile near the GC. Numerical simulations suggest that baryonic feedback processes can flatten the inner density distribution, potentially leading to a cored profile~\cite{Chan:2015tna}. To quantify the impact of this uncertainty, we assess how the presence of a central core affects the exclusion contour shown in the left panel of Fig.~\ref{fig:money_k}. Specifically, for each DM mass in the relevant range, we determine the core radius $r_c$ required to raise the exclusion parameter $\kappa(M_{\chi})$ to unity.

The procedure is as follows. We introduce a \textit{truncated} Einasto profile, defined as
\begin{equation}
    \rho_{\rm TEin}(r) = 
    \begin{cases}
        \rho_{\rm Ein}(r_c) &\qquad \mbox{for }  r < r_c \\
        \rho_{\rm Ein}(r) &\qquad \mbox{for }  r \geq r_c
    \end{cases} \ ,
\end{equation}
where $\rho_{\rm Ein}(r)$ is given in~\eqref{eq:Ein_profile}.  We then compute the $J$-factor using~\eqref{eq:flux}, with $\rho(r) = \rho_{\rm TEin}(r)$. Importantly, the modification of the central core must be performed under well-defined constraints to ensure physical consistency. As already discussed, the condition $\rho(r_{\odot})=0.4\,\text{GeV/cm}^3$ fixes $\rho_s$, while the free parameter $r_c$ is obtained by weakening the exclusion curve $\kappa(M_{\chi})$ to the threshold value $\bar{\kappa}=1$ for each DM mass 
\begin{equation}
    \begin{cases}
        \rho_{\rm TEin}(r_{\odot} = 8.5\,\text{kpc}) = 0.4\,\text{GeV/cm}^3
        \\[5pt]
        J_{\rm TEin}(\rho_s, r_s, r_c) \bar{\kappa} =  J_{\rm TEin}(\rho_s, r_s, 0)\kappa(M_{\chi}) \ .
    \end{cases}
\end{equation}
The second equation follows from the fact that rescaling the $\kappa$ parameter is mathematically equivalent to a direct rescaling of the $J$-factor. To anchor the analysis, we note that core radii $r_c \gtrsim 2\,\mathrm{kpc}$ (roughly the size of the Galactic bulge) are loosely disfavored by existing constraints~\cite{Hooper:2016ggc, Portail_2015}, and we adopt this value as a conservative lower bound on the DM density within our RoI. For such a core, we find $J = 3.6 \times 10^{22}~\si{GeV^2/cm^5}$, which is only a factor of 3 below the value reported in Table~\ref{tab:target_MW}, confirming that the uncertainty in the DM density profile within the selected RoI remains relatively well controlled.

 We repeat this procedure for each DM mass in the range $12\,\text{TeV} \le M_{\chi} \le 15\,\text{TeV}$ and show the resulting core radius required to achieve $\bar{\kappa} = 1$ as a function of $M_{\chi}$ in the right panel of Fig.~\ref{fig:money_k}. Focusing on the thermal mass window, we find that the resonant peak in the BS annihilation cross section at $M_{\chi} \simeq 13.5\,\text{TeV}$ leads to a robust exclusion, as the core size required to weaken the constraint to $\bar{\kappa} = 1$ exceeds by far $r_{\odot}$. Due to this resonant feature, we observe that for most masses below the central thermal value, a core significantly larger than the Galactic bulge~\cite{McMillan_2016,Portail_2015}) would still be needed to suppress the exclusion. Consequently, masses in the range $13.4\,\text{TeV} \lesssim M_{\chi} \lesssim 13.6\,\text{TeV}$ are robustly excluded by \textsc{Fermi-LAT} measurements of the Galactic diffuse emission. The situation changes for larger masses within the thermal window, where the core size required to weaken the constraint drops rapidly. In particular, for the central value $M_{\rm DM} = 13.7\,\text{TeV}$, we find a border-line exclusion, since any core size $r_c$ is sufficient to raise $\kappa$ above unity.

As discussed above in relation to the left panel of Fig.~\ref{fig:money_k}, when considering DM masses outside the thermal mass window, we observe a second peak around $M_{\chi} \simeq 12.2\,\text{TeV}$ with $r_c \simeq 7.5\,\text{kpc}$, reflecting the resonant feature of $\left<\sigma v\right>_{WW}$. Notably, due to this feature, the core size required to weaken the exclusion for masses in the range $12\,\text{TeV} \lesssim M_{\chi} \lesssim 13.4\,\text{TeV}$ remains overall larger than the size of the Galactic bulge, implying a robust exclusion. Finally, for DM masses above the thermal window, we find that  core sizes comparable to zero are sufficient to suppress the exclusion constraint, as the exclusion curve $\kappa(M_{\chi})$ lies already above $\bar\kappa$.

\subsection{CTAO Prospects}
\label{sec:CTAConstr}

In this section, we present the CTAO sensitivity to the MDM 5-plet. In particular, since observational data are not yet available, we can now present the prospects in terms of the observation time of dSphs needed to achieve a 95\% CL exclusion of the MDM 5-plet. The required observation time depends on the interplay between the low- and high-energy features of the spectrum, specifically the SE continuum in the hundred-GeV range and the photon line at the endpoint of the spectrum. After outlining the procedure for computing the DM and background event counts, we apply the statistical procedure to determine the observation time.
	
	\subsubsection{Statistical Analysis and Results}
	We consider $\mathcal{N}=20$ energy bins, each 0.2 decades wide, spanning the range from $25$ GeV to $160$ TeV. This binning is provided by CTAO together with the background event rate per solid angle of the the northern hemisphere, $\Delta \Gamma_{\rm bkg}/\Delta \Omega$, in the  website of~\cite{CTA:web}. Concerning the spatial features of our analysis, we focus on the RoI with angular aperture $\Delta\theta_{\rm dSphs}=0.5^{\circ}$ centered on the dSphs under observation, as suggested by the CTAO collaboration~\cite{CTAO:2024wvb}. Once the binning is provided, the number of background events in the $i$-th energy bin is
	\begin{equation}\label{eq:CTAbkg}
		N_{\rm bkg}^{i}=T_{\rm obs}\frac{\Delta\Gamma^i_{\rm bkg}}{\Delta\Omega}\Delta\Omega_{\rm dSphs}.
	\end{equation}
	Here $\Delta\Omega_{\rm dSphs}=4\pi\sin^2{\left(\Delta\theta_{\rm dSphs}/2\right)}\approx 2.4 \times 10^{-4}~$sr is the solid angle of the RoI of the considered dSphs and $T_{\rm obs}$ is the observation time. On the other hand, the number of events of the DM signal in the $i$-th bin reads
	\begin{equation}\label{eq:DMevents}
		\begin{split}
			&N_{\text{DM}}^{i}=T_{\rm obs}\int_{E_i}^{E_{i+1}}dE^\prime\,\,\frac{d\Gamma_{\text{DM}} (E^\prime)}{dE^\prime},\\
			&\frac{d\Gamma_{\text{DM}}(E^\prime)}{dE^\prime}=\int_{0}^{M_{\chi}}dE_{\gamma}\frac{d\Phi_{\text{DM}}(E_{\gamma})}{dE_{\gamma}}\mathcal{A}(E_{\gamma})R(E_{\gamma},E^\prime),
		\end{split}
	\end{equation}
	where $\mathcal{A}(E)$ is the CTAO effective area and the differential flux is obtained from~\eqref{eq:flux}. Notice that the first integral of~\eqref{eq:DMevents} runs over the \textit{reconstructed energy} $E^\prime$, while the second integral runs over the \textit{true energy} $E_{\gamma}$ of the prompt photons. To this purpose,  we model the energy resolution of the detector as a Gaussian   
	
	\begin{equation}\label{eq:res}
		R(E_{\gamma},E^\prime)=\frac{1}{\sqrt{2\pi\sigma(E_{\gamma})^2}}e^{-\frac{(E_{\gamma}-E^\prime)^2}{2\sigma(E_{\gamma})^2}},
	\end{equation}
	with $\sigma(E_{\gamma})\equiv\delta_{res}(E_{\gamma})E_{\gamma}$ and $\delta_{res}$ is provided in Ref.~\cite{CTA:web}.

	In order to determine the observation time $T_{\rm obs}$ required to exclude the candidate, we perform a statistical analysis by fixing the $\kappa$ parameter to unity. Also in this case, our likelihood is built by assuming a Poissonian distribution for the number of events in each energy bin. In particular,
	\begin{equation}\label{eq:like1}
		\mathcal{L}(T_{\rm obs})=\prod_{i=1}^{\mathcal{N}}\frac{N_{\text{th}}^{i}(T_{\rm obs})^{N_{\text{obs}}^{i}(T_{\rm obs})}}{N_{\text{\rm obs}}^{i}(T_{\rm obs})!}e^{-N_{\text{th}}^{i}(T_{\rm obs})}\equiv \prod_{i=1}^{\mathcal{N}}\mathcal{L}_{i}(T_{\rm obs}),
	\end{equation}
	where $N_{\text{th}}^{i}(T_{\rm obs})=N_{\text{DM}}^{i}(T_{\rm obs})+N_{\text{bkg}}^{i}(T_{\rm obs})$ is the theoretical number count, and we assume that the observed number of events $N_{\text{obs}}^{i}$ is given by the background \textit{only} hypothesis, i.e. $N_{\text{obs}}^{i}=N_{\text{bkg}}^{i}(T_{\rm obs})$. On top of~\eqref{eq:like1}, we add the contribution to the likelihood given by the systematic uncertainty on the $J$-factor. Precisely, we treat the $J$-factor as a nuisance parameter and marginalize over it~\cite{Lefranc:2016dgx,Lefranc:2016fgn}
	
	\begin{equation}
		\mathcal{L}_{sys}(\kappa)=\prod_{i=1}^{\mathcal{N}}\max_J \left[\mathcal{L}_{i}(T_{\rm obs})\times\mathcal{L}^J\right], \qquad \mbox{with } \mathcal{L}^J=\frac{1}{\ln (10) J_{\rm obs}}\mathcal{G}(\log_{10} J|\log_{10} J_{\rm obs})
	\end{equation}
	is a Gaussian likelihood centered at the observed $J$-factor, $\log_{10}J_{\rm obs}$, with standard deviation $\sigma_{\log_{10}J}$ equal to the uncertainty on  $\log_{10}J_{\rm obs}$, see Tab.~\ref{tab:target_MW} for the numerical values.
    
    Next, we define the test statistic as $\text{TS}(T_{\rm obs}) = -2\ln\left(\mathcal{L}_{\rm sys}(T_{\rm obs})/\mathcal{L}^0_{\rm sys}(T_{\rm obs})\right)$, with $\mathcal{L}^0_{\rm sys}$ being the \textit{null hypothesis} likelihood. Finally, we  determine the observation time $T_{\rm obs}$ such that the sensitivity reaches the MDM cross section, i.e we solve $\text{TS}(T_{\rm obs}) = 2.71$ for $T_{\rm obs}$ to obtain a 95\% CL limit on the observation time. Notice that the physical variable $T_{\rm obs}$ enters into both the signal and background event calculations, as evident from~\eqref{eq:CTAbkg} and~\eqref{eq:DMevents}. Therefore, neglecting systematics on the $J$-factor, we expect the sensitivity to the DM annihilation cross section to scale as the inverse square root of the observation time. 
    
	\begin{figure}[t!]  
		\centering\includegraphics[width=0.6\textwidth]{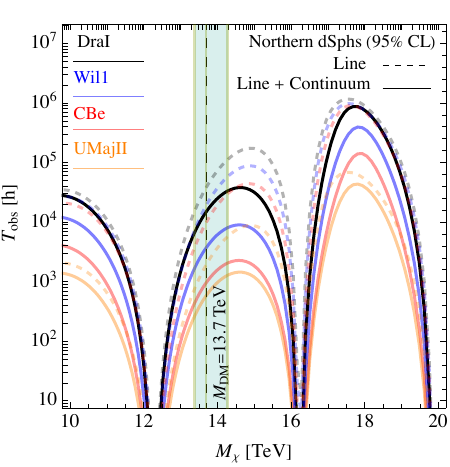}
		
\caption{
Observation time $T_{\text{obs}}$ required to derive a 95\% CL limit to the MDM 5-plet signal as a function of the DM mass for selected northern dSphs. Dashed lines represent the sensitivities obtained under the line-only signal hypothesis, while solid lines include both line and continuum contributions. The aquamarine band highlights the thermal mass window centered at the central value $M_{\rm DM} = 13.7\,\text{TeV}$. Results are presented for DraI, WilI, CBe, and UMajII.
}\label{fig:obsT} 
	\end{figure}

    \smallskip
	 We consider as our reference targets the sample dSphs in the northern hemisphere discussed in Tab.~\ref{tab:target_MW}. The results of the statistical analysis are shown in Fig.~\ref{fig:obsT}, where we present the observation time required to exclude the MDM 5-plet as a function of the DM mass in the range $10\,\text{TeV} \leq M_{\chi} \leq 20\,\text{TeV}$. The thermal mass window is indicated by a shaded aquamarine band. The solid lines in Fig.~\ref{fig:obsT} represent the results of the full spectrum (line+continuum). We obtain that the best sensitivity is achieved with observations of UMajII, for which we find that approximately $T^{\rm UMajII}_{\rm obs} \simeq 600$ hours are required to exclude the central mass value $M_{\rm DM} = 13.7\,\text{TeV}$. It is also worth noting that the required observation time varies significantly within the thermal mass range: we find $T^{\rm UMajII}_{\rm obs} \simeq 300$ hours at $M_{\rm DM} = 13.4\,\text{TeV}$ and $T^{\rm UMajII}_{\rm obs} \simeq 1300$ hours at $M_{\rm DM} = 14.3\,\text{TeV}$. As seen from Fig.~\ref{fig:annxsecs}, these sharp variations are a consequence of the resonant structure in the annihilation cross section, which induces an oscillatory behavior in the required observation time. Coming to the other ultrafaint dSphs, CBe and Wil1, we obtain weaker sensitivities due to the larger impact of systematics and the smaller values of the $J$-factors (see Tab.~\ref{tab:target_MW}). In particular, we find that $T^{\rm CBe}_{\rm obs}\sim 1000$ hours and $T^{\rm Wil1}_{\rm{obs}}\simeq 3700$ hours for $M_{\rm DM} = 13.7\,\text{TeV}$, while at the extreme of the thermal mass window we obtain $T^{\rm CBe}_{\rm{obs}}\simeq 500$ hours and $T^{\rm Wil1}_{\rm{obs}}\simeq 1800$ hours for $M_{\rm DM} = 13.4\,\text{TeV}$, and  $T^{\rm CBe}_{\rm{obs}}\simeq 2100$ hours and $T^{\rm Wil1}_{\rm{obs}}\simeq 8100$ hours for $M_{\rm DM} = 14.3\,\text{TeV}$. Finally, the weakest limits are obtained for the classical dSph DraI, for which we find $\sim\mathcal{O}(10^4)$ hours to exclude the thermal mass window. Although the resulting limits are weaker than those obtained from UMajII, they are based on a more robust and reliable determination of the $J$-factor.

    It is particularly interesting to notice how significantly the photon continuum contributes to the overall sensitivity. To illustrate this, we show with dashed lines in Fig.~\ref{fig:obsT} the observation time required when only the line component of the photon spectrum is included. For instance, in the case of DraI, we find that incorporating the continuum alongside the line reduces the required observation time by a factor of approximately $2.5$ compared to the line-only analysis, highlighting the significant role played by the continuum in enhancing the sensitivity relative to a line-only search. Similarly, for our best candidate, UMajII, we see that including the $\gamma$-ray continuum in the analysis improves the observation time obtained with a line-only analysis by a factor $\sim 3$. It is worth to notice that, when comparing the line-only and line+continuum results, the systematic uncertainties on the  $J$-factor have a greater impact on the line searches. This is because a spectral line is a sharp feature in the spectrum, affecting only a small number of energy bins. As a result, the systematic uncertainty from the $J$-factor leads to a stronger penalization of the sensitivity compared to continuum searches, where the deviation with respect to the background only hypothesis is distributed over a larger number of bins. As a consequence, we find that in the line-only case, the CTAO sensitivity across the various dSphs tends to be comparable, despite the differences in the central values of the $J$-factors, consistently with Fig.~8 of Ref.~\cite{CTAO:2024wvb}.

\section{Conclusion}\label{sec:conclusions}
Minimal Dark Matter represents the prototypical realization of the WIMP paradigm. Once the EW multiplet is fixed and a thermal freeze-out production mechanism is assumed, the Dark Matter phenomenology is entirely determined, making it a highly predictive and compelling candidate. Among all detection strategies, indirect detection via upcoming $\gamma$-ray astronomy offers the most decisive test: it will either lead to the discovery of, or definitively rule out, the smallest accidentally stable real representation, the Majorana SU(2) 5-plet. 

\smallskip
In this work, we have improved the existing literature among several directions. We have recomputed the Sommerfeld-enhancement (SE) affecting annihilation cross sections into all electroweak (EW) gauge bosons, including next-to-leading-order (NLO) corrections to the non-relativistic EW potential. In addition, we have incorporated the contribution from Dark Matter bound-state formation (BSF), analyzing its velocity dependence. At the practical level, NLO corrections have only a mild effect on the Sommerfeld-enhanced annihilation cross sections. However, they significantly impact the BSF contribution by shifting its peak cross section into the thermal mass window at $M_\chi \simeq 13.5~\si{TeV}$. We found that SE saturates at low relative velocities, $\vrel \lesssim 10^{-2}$, while, the cross section for formation of the dominant $^1(1s)_3$ bound state is p-wave suppressed in low-velocity environments such as dwarf spheroidal galaxies (dSphs), while its contribution becomes significant in the Galactic halo. 

\smallskip
Including such non-perturbative effects leads to distinctive and unique features in the resulting $\gamma$-ray spectrum. In particular, Sommerfeld enhanced annihilation into heavy EW gauge bosons produces a smooth continuum component extending from a few hundred GeV up to the few-TeV range, while direct annihilation into hard photons yields a prominent spectral line centered at the Dark Matter mass. For the latter feature, we have included resummation effects at next-to-leading-log (NLL) accuracy, computed by following SCET formalism providing the most precise prediction of the spectral shape to date. Moreover, we have computed for the first time the spectral features arising from BSF, which result in a spectral line centered at the typical binding energy $E_b\simeq73~\si{GeV}$, accompanied by a sizable continuum contribution in the hundreds of GeV range. With the full spectrum in hand, we have evaluated the differential $\gamma$-ray flux and explored potential indirect detection avenues for probing the MDM 5-plet. Our analysis is structured in two parts: low-energy constraints based on \textsc{Fermi}-LAT observations of Galactic diffuse emission, and high-energy projections using the forthcoming Cherenkov Telescope Array Observatory (CTAO), which will probe the multi-TeV regime with unprecedented energy resolution.

\smallskip
Focusing on the inner Galaxy, we update previous constraint on the Galactic diffuse emission, analyzing 16 years of \textsc{Fermi-LAT} data. We find that \textsc{Fermi-LAT} already enable robust exclusion of the lower edge of the thermal mass window for the MDM 5-plet. This constraint stems from the proximity of these masses to the resonant peak of the BSF cross section, which amplifies the signal via both the spectral line at the typical binding energy $E_b\simeq 73~\si{GeV}$ and the associated continuum component. For this case, we have shown that core sizes larger than the Galactic bulge would be required to evade exclusion. In contrast, for masses closer to the central or upper edge of the thermal mass window, the exclusion becomes less robust due to their distance from resonant features in the annihilation cross sections. For these cases, core sizes smaller than bulge radius generally suffice to evade the constraints. Shortly before the submission of our work, a closely related study appeared in the literature~\cite{Safdi:2025sfs}. While our results are generally consistent with theirs, we note some discrepancies that have an important impact in the interpretation of the results, particularly in the right edge part of the thermal mass window. We find a discrepancy in the continuum component of the $\gamma$-ray spectrum. This originates from not considering NLL resummation, which significantly impacts the $Z$-boson contribution, as highlighted in~\eqref{eq:C_spectrum}. Neglecting this effect inevitably leads to an overestimation of the SE annihilation cross section. Additionally, despite the use of the same method to compute the BSF cross section, their result appears to overestimate it by a factor of 2. Moreover, although they implement NLO corrections to compute the SE annihilation cross sections, they do not consistently update the thermal mass window, instead using an older symmetric range centered at 13.6 TeV~\cite{Bottaro:2021snn}. As a result of these considerations, the limits they derive at the right edge of the thermal mass window appear stronger than those obtained from our analysis.  

\smallskip
In contrast the high-energy part of the spectrum, only mildly affected by BSF, serves to forecast the required observation time for several of the MW's dSphs using CTAO. We stress that the northern site in La Palma will become operational in the coming years, earlier than the southern array, providing an optimal target for the observation of dSphs. In this context, we find that including the $\gamma$-ray continuum from Sommerfeld-enhanced annihilation into gauge bosons significantly extends the reach within the thermal mass window compared to a line-only analysis. Notably, we find that approximately 600 hours of observation of the ultra-faint dSph Ursa Major II would suffice to probe the central mass value of the MDM 5-plet. It is important to emphasize, however, that the $J$-factor for ultra-faint dSphs remains subject to improvement as more precise kinematical data become available.

\smallskip
In conclusion, prior to the advent of the next generation of kiloton-scale direct detection experiments, $\gamma$-ray indirect detection remains the most promising strategy, capable of exploiting current data and forthcoming searches to potentially probe the pure electroweak nature of Dark Matter.

\section{Acknowledgments}
We would like to thank Salvatore Bottaro for his insightful discussions. Although not listed as an author, we gratefully acknowledge his significant input, which substantially influenced the development of this study. We also thank  Giovanni Armando and Daniele Gaggero for their involvement in the early stages of the project. Our work further benefited from valuable discussions with Michele Doro, Christopher Eckner, Nicholas L. Rodd, Tracy R. Slatyer and Gabrijela Zaharijas. This work receives partial funding from the European Union–Next generation EU (through Progetti di Ricerca di Interesse Nazionale (PRIN) Grant No. 202289JEW4).

\bibliographystyle{JHEP}
{\footnotesize
\bibliography{main}}

\end{document}